\documentclass[
  ,draft            
  ]
  {aipproc}

\layoutstyle{6x9}

\begin{document}

\title{The decade of galaxy formation: pitfalls in the path ahead}


\author{Simon P. Driver}{address={School of Physics, University of St Andrews, St Andrews, Fife, KY16 9SS}}



\begin{abstract}
At the turn of the decade we arguably move from the era of precision
cosmology to the era of galaxy formation. One approach to this problem
will be via the construction of comprehensive galaxy samples. In
this review I take the opportunity to highlight a number of challenges
which must be overcome before we can use such data to construct a
robust empirical blueprint of galaxy evolution. The issues briefly
highlighted here are: the Hubble tuning fork versus galaxy components,
the hierarchy of structure, the accuracy of structural decompositions,
galaxy photometry, incompleteness, cosmic variance, photometric versus
spectroscopic redshifts, wavelength bias, dust attenuation, and the
disconnect with theory. These concerns essentially form one of the key
motivations of the GAMA survey which, as one of its goals, will
establish a complete comprehensive kpc-resolution 3D multiwavelength
(UV-Opt-IR-Radio) database of 250k galaxy systems to $z <0.5$.
\end{abstract}

\maketitle


\section{Introduction}
Following a decade of precision cosmology we move confidently into the
decade in which we expect the process(es) of galaxy formation to become
clear. The problem is well defined, how did baryonic matter assemble
from the smooth distribution of neutral atoms at recombination to the
$\sim 10^9 $M$_{\odot}$ conglomerations of dust, gas, and stars that
we see today. As we enter 2010 it is sobering to have to acknowledge
that we do not yet have a cohesive model of galaxy formation
\cite{pn10} but rather an extensive body of empirical results from
diverse approaches and which are often in conflict (does the IMF vary
or not?, why does the cosmic star-formation history overpredict the
local stellar mass density?, is the merger rate high or low? how
exactly do AGN fit in?), and that these fragmented bodies of evidence
sit uncomfortably with the obvious bottom-up merger-rich inference one
naturally draws from the hierarchical-CDM paradigm (why does the
majority of stellar mass sit in thin fragile discs if merging is the
principle formation mechanism?  why do both galaxies and AGN follow
downsizing trends if the majority of massive systems form late?).

At the outset of this journey it may be worth lightening our burden in
two ways: firstly by acknowledging that probably a significant
fraction of the current literature is in error (or at least dominated
by poorly quantified systematics), and secondly that while
hierarchical-CDM has been spectacularly successful at explaining the
large-scale structure ($>1$Mpc), the exact processes by which the
baryons evolve remains a matter of prescriptive
speculation. Monolithic collapse, feedback, merging, and hot/cold gas
infall sure, but how much of each, and how are they regulated? The
modeling, while a vital industry and visually spectacular, should not
overshadow or overly influence the empirically driven process of
discovery, but rather augment it. Nor should we be over-burdened with
statistical precision given the domination of the data by hidden
systematics. So many of the papers I referee focus exquisitely on the
statistical errors while paying scant regard to the systematics
errors, some of which are laid out here.

Despite the plight of the world economies, we move into the new decade
with a truly outstanding set of facilities coming online, for example:
the refurbished HST, the HERSCHEL and WISE missions, the commissioning
of VISTA and ALMA, and the opening up of 21cm studies beyond 12000km/s
via ASKAP and MeerKAT, and in the longer term JWST, LSST, the SKA, and
SNAP/DUNE/EUCLID (and too many more to mention). It is up to us not to
fumble the opportunity this problem and these facilities provide, nor
be overly fettered by the preconceptions we carry with us.

In this review I highlight ten practical issues which together
convince me that low rather than high redshift is the place to start,
that the time has come to truly embrace the multiwavelength approach,
and that as much as we enjoy working in small teams the way forward is
through large, open, and internationally cooperative collaborations.

\section{Challenges in the road ahead}

\subsection{Global versus component measurements}
What exactly constitutes a galaxy? Is it a single entity best defined
by global parameters (size, colour, flux), or a more complex
construction of distinct components (bulge, bar, disc)? Which
components are transient and which might represent a fundamental
imprint of its formation history? 

The legacy of Hubble's initial classification scheme devised over 80
years ago lives on as galaxies are most often defined by their Hubble
type (a global measure), but why? If the bulge and disc are co-eval
entities whose properties are co-dependent then this makes sense but,
putting aside the question of pseudo-bulges, it is difficult to find
any property beyond total luminosity which discs and bulges share. The
ages, metallicities, dynamics, dust, star-formation rates are all
distinct evoking the notion of a least two fully distinct evolutionary
processes and pathways. A classical bulge shares more commonality with
a giant elliptical than the disc so surely it is logical to isolate
the bulges and discs and look for commonalities and correlations
within each structural group rather than with Hubble type. Perhaps this is
best illustrated through galaxy bimodality (see Fig. 1), where the
blue cloud and red sequence (global measurements) provide a less clear
distinction than that provided by the bulges and discs (component
measurements). As a community we know this deep down but the blue/red
mantra along with the Hubble Tuning fork is proving hard to leave
behind.

\begin{figure}
  \includegraphics[height=0.5\textwidth]{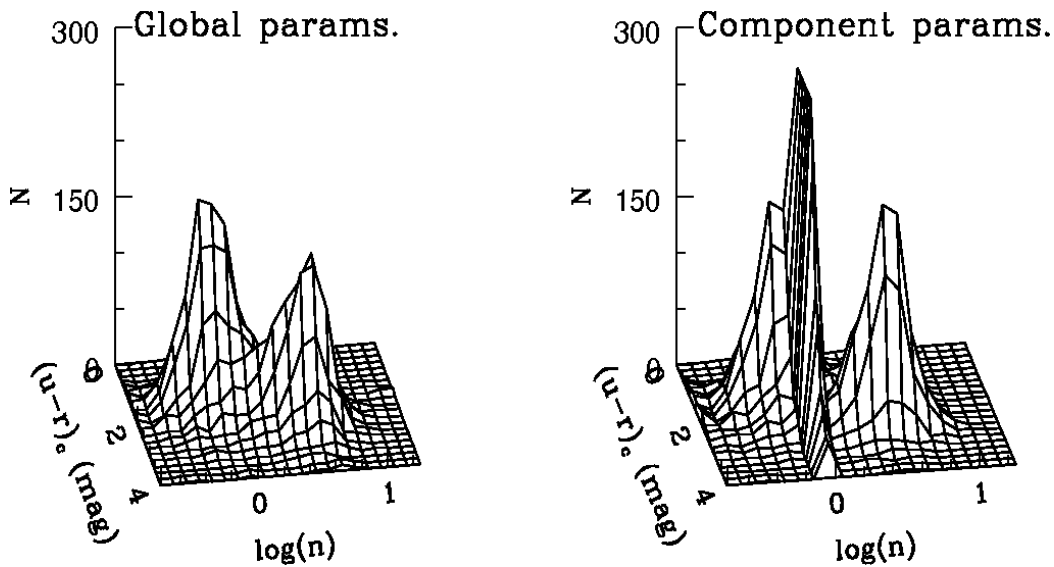}

\caption{The colour versus S\'ersic index plane for (left) global
  parameters and (right) component parameters. The division appears
  stronger and cleaner for the component params with little sign of
  any ``green valley'' indicating this is a more fundamental
  division.}
\end{figure}

\subsection{The hierarchy of components}
The spatial distribution reflects the combined orbits of billions of
stars which are not trivially established or perturbed. This is
distinct to spectroscopic information which contains a rich chemical
imprint but only upto the last major episode of star-formation
(typically a few Gyrs). While dynamics are robust to star-formation
(except in the case of a rare major-merger), composite spectra are not
and so the spatial distribution is more likely to contain an imprint
of the full lifetime of the system and spectroscopy the current
state-of-play. However despite the potential of structure to probe the
full dynamical history the (detailed) study of galaxy structure is
currently limited to the very local domain and mostly in atypical
cluster environments (e.g., Virgo, Fornax, and Coma).

If one accepts that structure (and by this one means the stable and
dynamically coupled configurations of billions of stars) is a
byproduct of the formation history the next key question is whether
all structure is important or whether their exists a hierarchy. For
example one could attempt to separate galaxies into nucleus, bulge,
pseudo-bulge, bar, inner disc, outer disc, outer truncation but this
might be going too far. Fig.~2 shows two possible hierarchies
(schemes) and readers will probably identify others. This is a key
area where the theorists and simulators can inform the observers as to
which structures are fundamental, which are transient, and which
measurable quantity might best connect to key physical processes
(mechanism, timescales and rates).

\begin{figure}
  \includegraphics[height=0.75\textwidth,angle=90]{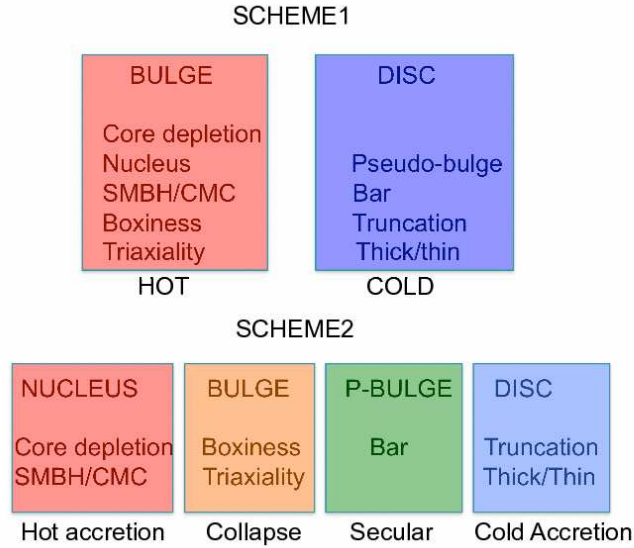}
  \caption{Which structures are key and how do they relate to
    formation processes. Two schemes are shown one which identifies
    the bulge and disc as the primary components along with secondary
    features and a second which place all components onto an equal
    footing.}
\end{figure}

\subsection{Structural decomposition}
One of the reasons structural decomposition has stalled at low
redshift is the difficulty in unambiguously profiling galaxies as
their resolution dwindles \cite{b10} and in the presence of dust
\cite{g10}. From the ground credible bulge-disc decomposition is
probably limited to $z<0.1$ where the spatial resolution approaches
typically 1kpc per arcsecond. With HST one can actually resolve bound
structures to $<1$kpc resolution across the full path-length of the
Universe thanks to the diameter-distance relation but one is of course
hit by the surface brightness dimming, the shift to more clumpy
shorter rest-wavelengths, and perhaps more fundamentally the apparent
evolution in structure from grand-design (order) at low redshift to
train-wrecks (chaos) at high redshift. Exactly when ordered structure
appears and how this depends on environment are key questions we
should be able to address if we remain mindful of the selection biases
at work. Fig.~3 shows two equally valid decompositions for a nearby
galaxy, one with a bar and one without, but which is correct and how
to robustly automate the decision making process. Some will argue,
with good reason, that dynamical information will always be required
to unambiguously identify bar and pseudo-bulge components.

\begin{figure}
  \includegraphics[height=0.35\textwidth]{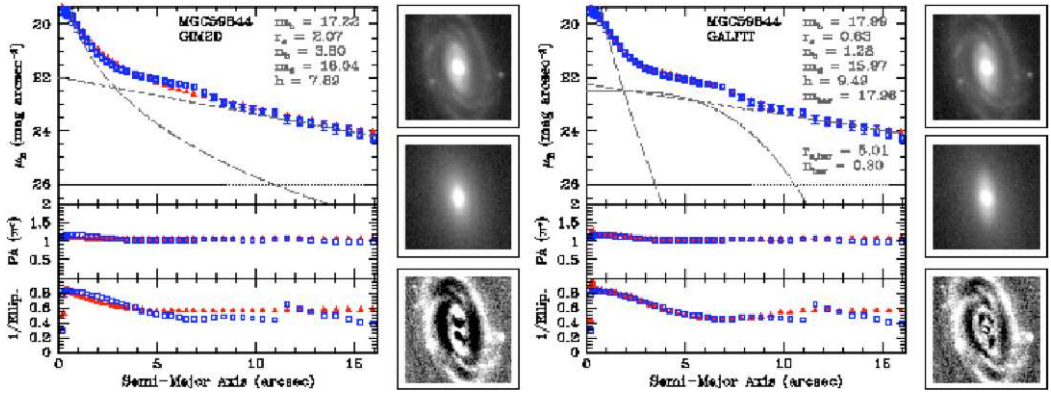}
  \caption{Two structural decompositions of the same galaxy, one
    without (left) and one with (right) a bar, but which is correct?
    Figure adapted from Cameron \& Driver (2009).}
\end{figure}

\subsection{Photometry}
Even the basic question of exactly how to measure the total flux of a
galaxy is unclear. In common usage are circular Petrosian (SDSS;
UKIDSS), elliptical Kron (MGC; SExtractor) and circularised elliptical
isophotal (2MASS). However none of these will actually be meaningful
if only the very central portion ($<1$R$_e$) of the galaxy was
initially detected (e.g., if one detects just the bulge none of these
methods will recover the disc). For example it is often stated that an
SDSS Petrosian magnitude will recover 80\% of the flux of an ($n=4$)
elliptical galaxy. However this is only correct if the true intrinsic
effective radius is sampled in the initial isophotal detection process
(rarely the case for high-z or low luminosity local systems)
\cite{gd05}. A promising way forward is profile or S\'ersic fitting
via sophisticated image-analysis packages such as GALFIT3, GIM2D and
BUDDHA, however one still has to deal with what lies beneath the
isophote in terms of truncation, and anti-truncation of the outer disc
profile \cite{pt06}. As a consequence it is quite easy to
underestimate both fluxes (and sizes) of low surface brightness (i.e.,
low luminosity and high-z objects) by a factor of 1mag or more. The
question once posed by Mike Disney \cite{d76} as to whether we are
missing populations of giant galaxies (aka crouching giants) has
perhaps morphed into a much more subtle question as to how much light
are we missing from the galaxies we detect. The only credible solution
appears to be deeper imaging yet this invariably reveals more
structural complexity and extreme asymmetry (see discussion of faint
outer structures in the review by Annette Ferguson in these
proceedings). Fig.~4 highlights the flux bias by showing S\'ersic
photometry minus SDSS photometry as a logarithmic function of the
S\'ersic index. As we see for high S\'ersic indices the photometric
error becomes severe even at low redshift \cite{gd05b}.

\begin{figure}
  \includegraphics[height=0.4\textwidth]{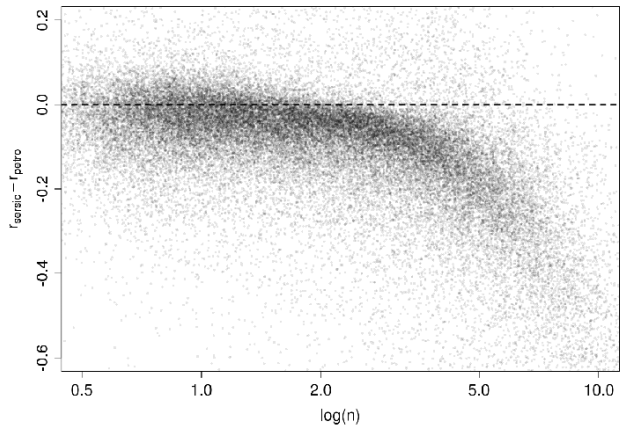}
  \caption{A diagram showing the systematic biases in flux measurement
    as a function of profile shape for bright nearby galaxies from the
    SDSS.}
\end{figure}

\subsection{Incomplete sampling}
Incompleteness can be defined in many ways, typically we use it to
refer to spectroscopic incompleteness whereby only some percentage of
an input catalogue has confirmed redshifts (e.g., 88\% 2dFGRS, 90\%
SDSS, 98\% MGC). Surely the missing few per cent doesn't matter?
Well...it depends what they are! If the missing systems are
preferentially low luminosity then the problem becomes serious: a tiny
population which is only observable within a tiny volume implies a
massive population with a massive error \cite{dp96}.  We also have
incompleteness in our input catalogue which can have two causes, (1) a
population of galaxies never detected, and/or, (2) a population whose
flux was systematically underestimated and therefore underrepresented
after any magnitude cut (i.e., missing galaxies and missing
light). The first of these is unlikely to affect the giant galaxies
where the majority of the total cosmic stellar mass lies \cite{d07},
however the second is an insidious bias which potentially effects
every galaxy \cite{cd09}. How severe is it?  Until we work out how to
measure galaxy photometry correctly this issue remains unclear
\cite{cd02}.  Ultimately these issues can only be minimised (and never
entirely overcome) by deeper imaging and higher spectroscopic
completeness both of which are not easily appreciated by time
allocation committees. Fig.~5 illustrates the parameter space
currently probed by the widest and deepest surveys implying our census
even locally is woefully incomplete.

\begin{figure}
  \includegraphics[height=0.5\textwidth]{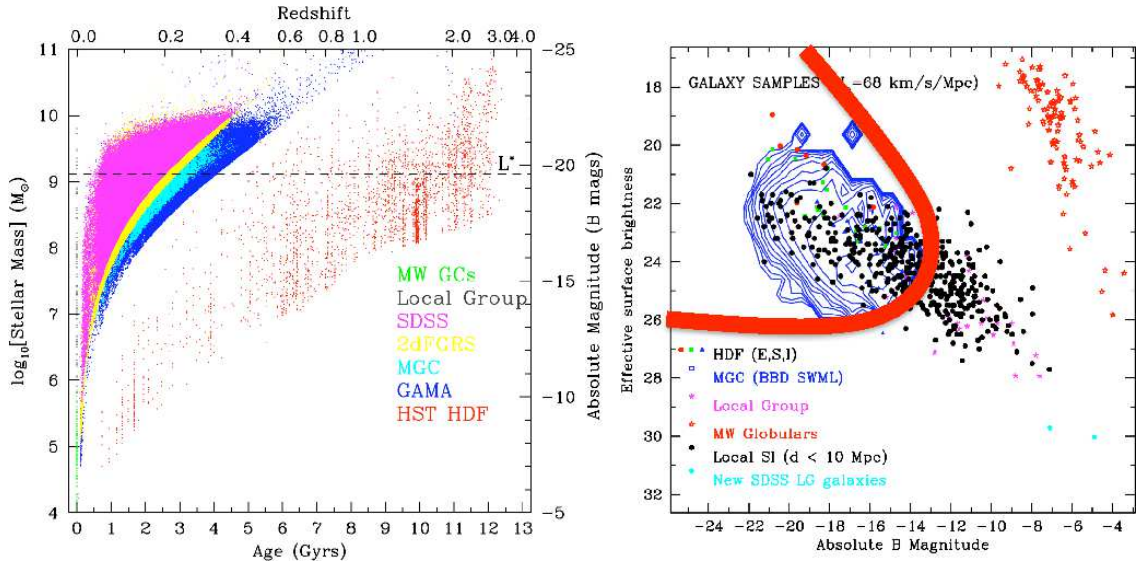}
  \caption{(Left) the mass-age plane showing which portions have been
    surveyed over the observable universe and (right) the sampling of
    the luminosity-surface brightness plans at redshift zero. In both
    cases there is far more blank space than comprehensively sampled
    parameter space implying much more work to be done even at low
    redshift.}
\end{figure}

\subsection{Cosmic Variance}
Cosmic (or sample) variance is present in all our surveys
\cite{dr10}. In fact one of the conclusions, buried in one of the
2dFGRS papers, is that deriving the power-spectrum for larger samples
is to some extent fruitless as it is perfectly plausible that our
entire observable universe could be a very slightly anomalous region
of the entire Universe. The exact scale at which the universe becomes
homogeneous is actually ill defined with interesting anomalies arising
in WMAP5. If the entire observable universe is potentially prone to
Cosmic Variance then your local or pencil beam survey is most
definitely prone to cosmic variance. Fig.~6 shows how cosmic variance
depends on total survey volume derived by slicing and dicing the
SDSS. Cosmic Variance only falls below 10\% once a volume of
$10^7$Mpc$^3$ has been sampled ($h=0.7$).

\begin{figure}
  \includegraphics[height=0.35\textwidth]{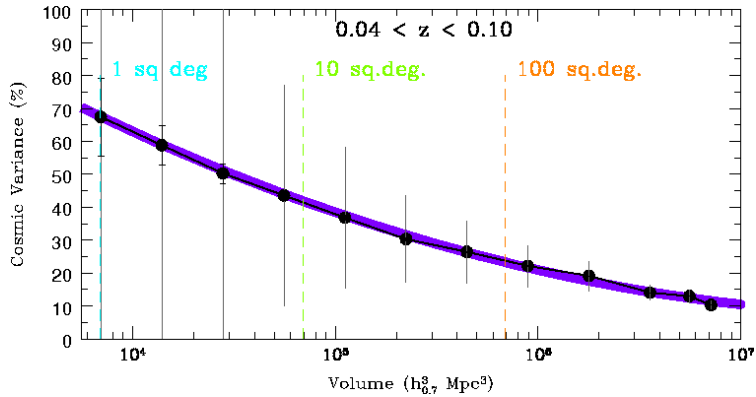}
  \caption{Cosmic variance (\%) as a function of survey volume.}
\end{figure}

\subsection{Photo-z versus spectro-z}
Fig.~7 shows cone plots from the GAMA survey using either $ugriz$
photometric redshifts (upper) or spectroscopic redshifts (lower). As
we probe to fainter fluxes we sample more actively star-forming
galaxies (flatter SEDs), the low luminosity population (intrinsically
flatter SEDs), and our SED measurements become noisier, as well as
more severely dust attenuated (see below). The photo-z method relies
heavily on the existence of a strong 4000\AA~break which will diminish
with limiting flux for the reasons indicated above. While photo-z's
allow us to move forward the resulting luminosity distributions and
various densities can only be taken as indicative at best.

\begin{figure}
  \includegraphics[height=0.75\textwidth]{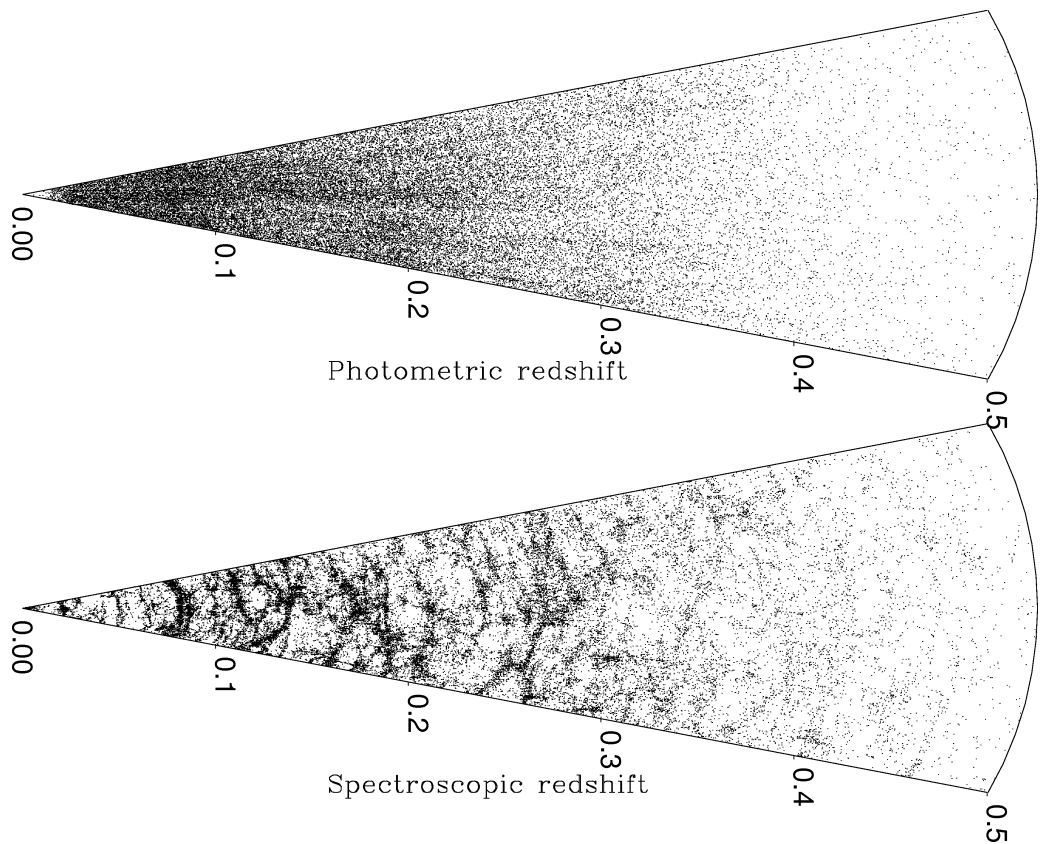}
  \caption{The GAMA 12hr region according to photo-z measurements from
    SDSS (upper) or spectroscopic measurements from the AAT (lower)}.
\end{figure}

\subsection{Wavelength Bias}
Fig.~8 shows the classic galaxy NGC891 viewed from the FUV through the
optical and into the near-IR. The star-formation lane broadens into
the stellar population with a dust lane that gradually fades and then
brightens as one sees only the glowing dust. The HI disc extends 5-10
times further than the stellar disc (see review by Lister
Staveley-Smith and others in these proceedings), and often engulfs or
shows HI bridges (streams) with otherwise apparently optically
detached systems. As galaxy formation is encapsulated in the
gas-star-dust cycle observations in the UV, optical, far-IR and radio
are an absolute necessity if one is to untangle this interplay.

\begin{figure}
  \includegraphics[height=0.5\textwidth]{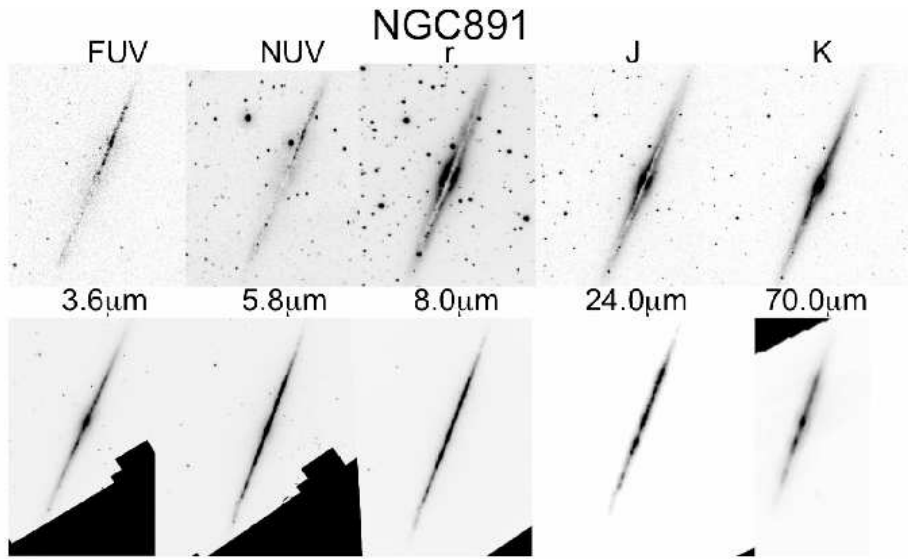}
  \caption{NGC891 observed at various wavelengths from UV to far-IR. Note how its appearance changes significantly with wavelength.}
\end{figure}

\subsection{Dust Attenuation}
For most of the past two decades we've managed to sweep dust under the
carpet. Having agreed to disagree in the 90s following the bloody-nose
era of optically thick and thin sandwiches and slabs \cite{ddp89}.
We're now seeing a multitude of results from the MGC
\cite{d07b},\cite{d08} and SDSS \cite{s07}, \cite{c07}, \cite{ur08},
\cite{ps08}, \cite{cp09}, \cite{m09}, \cite{g09}, \cite{m10},
\cite{y10} categorically demonstrating that the distribution of
face-on systems bears little resemblance to edge-on systems with
attenuation being measured in magnitudes rather than tenths of a
magnitude. Unless we wish to dispense with the CP we are left to
conclude that dust is playing havoc with our optical measurements and
at this moment in time represents by far the largest systematic
affecting all optical surveys. The problem gets worse towards higher
redshift as one views more heavily attenuated wavelengths and the
evolution of dust is unclear. Ultimately the D(ust)-correction may
actually be more severe than the better known E- and K-
corrections. To convey a scale of the problem our data demonstrate
unequivicably that only 50\% of the energy produced by stars actually
makes it out of the host galaxy. While we can quantify the mean
attenuation and its dependence on inclination the variance within the
galaxy population and how this depends on environment and structural
properties remains poorly quantified (although see \cite{c10}. In
effect, dust attenuation acts as a broad (many mag) smoothing filter
over all our distributions, moving edge-on giants into the same
luminosity bins as face-on dwarfs etc. There are two solutions here,
sophisticated SED modeling incorporating far-IR observations, or
simply start from square one in the near-IR (aka UKIDSS and VISTA).
Fig.~9 shows the impact of dust attenuation as a function of
inclination on the luminosity distributions of discs and bulges at
various wavelengths, indicating the severity of the effect.

\begin{figure}
  \includegraphics[height=0.6\textwidth]{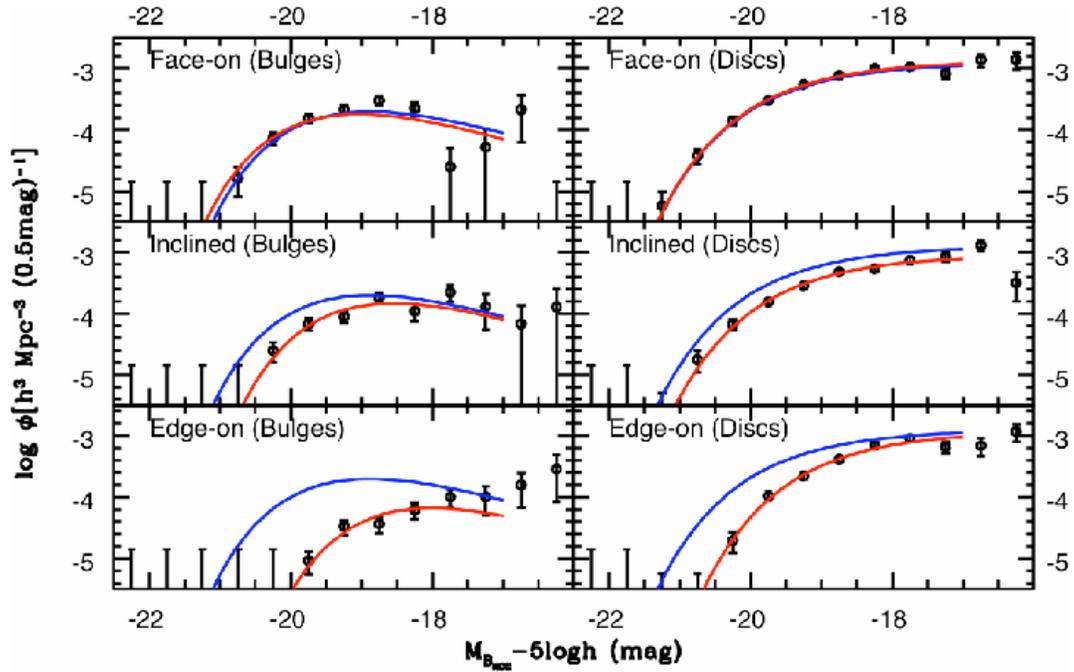}
  \caption{The impact of dust attenuation on the measurement of the
    bulge (left) and disc (right) luminosity functions versus hist
    galaxy inclination.}
\end{figure}

\subsection{Disconnect with hierarchical-CDM}
The disc of the Milky-way is thin, remarkably thin, and appears to
have formed some time ago and exhibited a remarkably quiescent
history. While one may argue that the Milky Way is a 10-sigma event it
certainly doesn't appear to be in any global way. So how in a
hierarchical-CDM Universe can such fragile structure come about and remain so
pristine for so long. Our recent census \cite{d08} indicates that
approximately 60\% of the stellar mass is in the form of extended
discs. If discs can only form after the last major merger event then
major mergers are the secondary formation mechanism confined to the
high-z Universe. The question we're left with is what is the dominant
mechanism?

\subsection*{The star-gas conspiracy}
Perhaps one of the most puzzling aspects of the empirical dataset,
which highlights the very different picture one might draw from
optical versus radio observations, and hence the need to go truly
multi-wavelength, is the comparison of the cosmic star-formation
history with the cosmic HI gas history as shown on Fig.~10. While the
cosmic star-formation shows a varied history the neutral gas content
of galaxies appears to stay constant. If borne out then one needs cold
gas to infall at precisely the rate require to counteract the gas lost
to star-formation. This suggests a remarkable conspiracy between the
stars and the gas but more importantly the need to better couple
optical and radio observations.

\begin{figure}
  \includegraphics[height=0.5\textwidth]{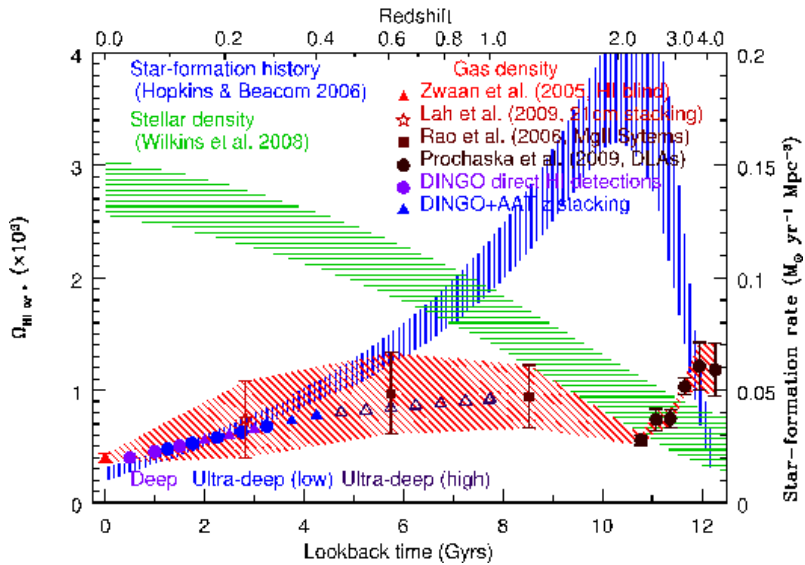}
  \caption{The cosmic star-formation history (blue), the build-up of
    stellar mass (green) and the cosmic HI history (red) which despite
    the intense star-formation history appears to stay constant.}
\end{figure}

~

My belief is that the answers to the above challenges do not lie in
the numerical simulations but in the empirical datasets heading our
way, and it is imperative that the people resources and appropriate
techniques are put in place to manage and merge these data
streams. Fig. 11 shows a possible starting point for our empirical
blueprint, showing the era of AGN activity which (because of the SMBH
connection) must link with bulge formation, and the era of
star-formation which must link with disc growth. As this dwindles we
perhaps see three clear phases: bulge growth, disc growth and the
current era of secular evolution. The first goal of the new decade
will be to attempt to ask whether the new data confirms or refutes
this simple sketch.

\begin{figure}
  \includegraphics[height=0.5\textwidth]{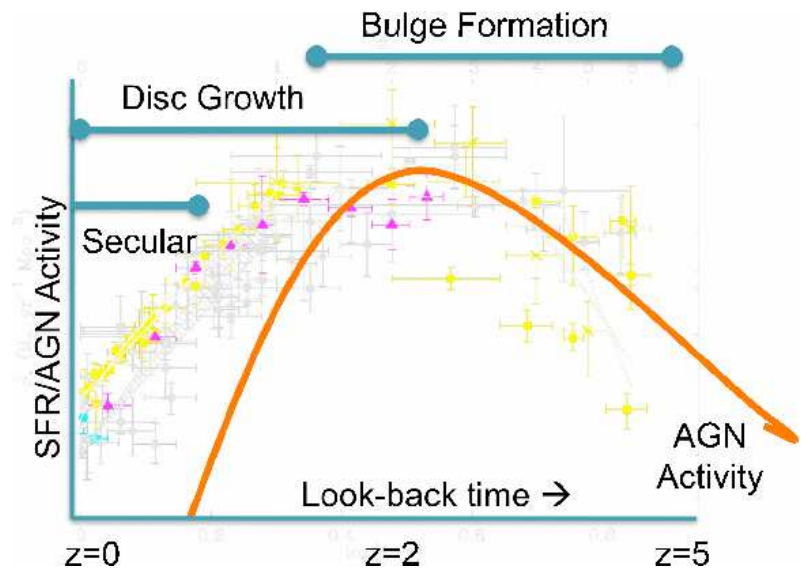}
  \caption{The start of an empirically determined galaxy
    formation blueprint showing three distinct evolutionary modes and
    eras?}
\end{figure}


\section{Galaxy And Mass Assembly (GAMA)}
From the above I argue that a new kind of galaxy database is needed
which is deep enough to provide complete and accurate photometry, has
sufficient resolution to enable bulge-disc decomposition, directly
samples the dust, stars and gas (opt, IR, radio), and with high
spectroscopic completeness and sample size as to squeeze out the
biases. The Galaxy And Mass Assembly (GAMA) survey \cite{d09} has been
designed with this in mind, and aims to survey a common block of sky
with GALEX, VST, VISTA, HERSCHEL, (WISE), and ASKAP while obtaining
spectroscopic redshift confirmation to $r<19.8$ mag with exceptional
completeness. The survey is nearing the completion of its initial AAT
allocation (66 nights) enabling the acquisition of 120k
systems. Incoming data from all of the imaging facilities will arrive
in 2010 followed by ASKAP from 2012, providing the best possible
database from which to start building a detailed empirical blueprint
of galaxy formation. More details on the GAMA project can be found in
Driver et al (2009) \cite{d09} or simply follow our progress on
astro-ph. Anyone wishing to become involved in the GAMA project should
feel free to contact spd3\@st-and.ac.uk.



\begin{theacknowledgments}
Thanks to the conference organisers for a great meeting and
collaborators new and old for many interesting discussions.
\end{theacknowledgments}



\bibliographystyle{aipproc}   

\end{document}